\documentclass[letterpaper]{article} 
\usepackage{aaai2026}  
\usepackage{times}  
\usepackage{helvet}  
\usepackage{courier}  
\usepackage[hyphens]{url}  
\usepackage{graphicx} 
\usepackage{natbib}  
\usepackage{caption} 
\usepackage{algorithm}
\usepackage{algorithmic}

\usepackage{array} 
\usepackage{booktabs} 
\usepackage{amsmath} 
\usepackage{multirow} 
\usepackage{amssymb}
\usepackage[cal=boondox]{mathalfa} 

\usepackage{newfloat}
\usepackage{listings}
\DeclareCaptionStyle{ruled}{labelfont=normalfont,labelsep=colon,strut=off} 
\floatstyle{ruled}
\newfloat{listing}{tb}{lst}{}
\floatname{listing}{Listing}
\author{
    Written by AAAI Press Staff\textsuperscript{\rm 1}\thanks{With help from the AAAI Publications Committee.}\\
    AAAI Style Contributions by Pater Patel Schneider,
    Sunil Issar,\\
    J. Scott Penberthy,
    George Ferguson,
    Hans Guesgen,
    Francisco Cruz\equalcontrib,
    Marc Pujol-Gonzalez\equalcontrib
}
\affiliations{
    \textsuperscript{\rm 1}The University of QueensLand\\
    \textsuperscript{\rm 2}University of Southampton\\
    \textsuperscript{\rm 3}Soochow University\\ 
    \textsuperscript{\rm 4}Chengdu Jiaozi Financial Holding Group Co., Ltd\\
    \textsuperscript{\rm 5}National University of Singapore\\
    \textsuperscript{\rm 6}The University of Hong Kong\\
    
    hongli.chen@uq.edu.au, p.fang@soton.ac.uk, craveuchen@gmail.com, e1664832@u.nus.edu, jinghao@connect.hku.hk, f.tang1@uq.edu.au, x.cai@soton.ac.uk, ssshan@suda.edu.cn, feng@eecs.uq.edu.au

}

\title{HiFi-Mamba: Dual-Stream $\mathcal{w}$-Laplacian Enhanced Mamba for High-Fidelity MRI Reconstruction}

\author {
    Hongli Chen\textsuperscript{\rm 1, \equalcontrib},
    Pengcheng Fang\textsuperscript{\rm 2, \equalcontrib},
    Yuxia Chen\textsuperscript{\rm 4},
    Yingxuan Ren\textsuperscript{\rm 5},
    Jing Hao\textsuperscript{\rm 6},
    Fangfang Tang\textsuperscript{\rm 1},
    Xiaohao Cai\textsuperscript{\rm 2},
    Shanshan Shan\textsuperscript{\rm 3, \thanks{Corresponding author.}},
    Feng Liu\textsuperscript{\rm 1}
}

\usepackage{bibentry}

\begin{document}

\maketitle

\begin{abstract}
Reconstructing high-fidelity MR images from undersampled k-space data remains a challenging problem in MRI. While Mamba variants for vision tasks offer promising long-range modeling capabilities with linear-time complexity, their direct application to MRI reconstruction inherits two key limitations: (1) insensitivity to high-frequency anatomical details; and (2) reliance on redundant multi-directional scanning. To address these limitations, we introduce High-Fidelity Mamba (HiFi-Mamba), a novel dual-stream Mamba-based architecture comprising stacked $\mathcal{w}$-Laplacian (WL) and HiFi-Mamba blocks. Specifically, the WL block performs fidelity-preserving spectral decoupling, producing complementary low- and high-frequency streams. This separation enables the HiFi-Mamba block to focus on low-frequency structures, enhancing global feature modeling. Concurrently, the HiFi-Mamba block selectively integrates high-frequency features through adaptive state-space modulation, preserving comprehensive spectral details. To eliminate the scanning redundancy, the HiFi-Mamba block adopts a streamlined unidirectional traversal strategy that preserves long-range modeling capability with improved computational efficiency. Extensive experiments on standard MRI reconstruction benchmarks demonstrate that HiFi-Mamba consistently outperforms state-of-the-art CNN-based, Transformer-based, and other Mamba-based models in reconstruction accuracy while maintaining a compact and efficient model design.


\end{abstract}

\begin{links}\link{Code}{https://github.com/PengchengFang-cs/HiFi-Mamba}
\end{links}

\begin{figure}[!t]
\centering
\includegraphics[width=1\columnwidth]{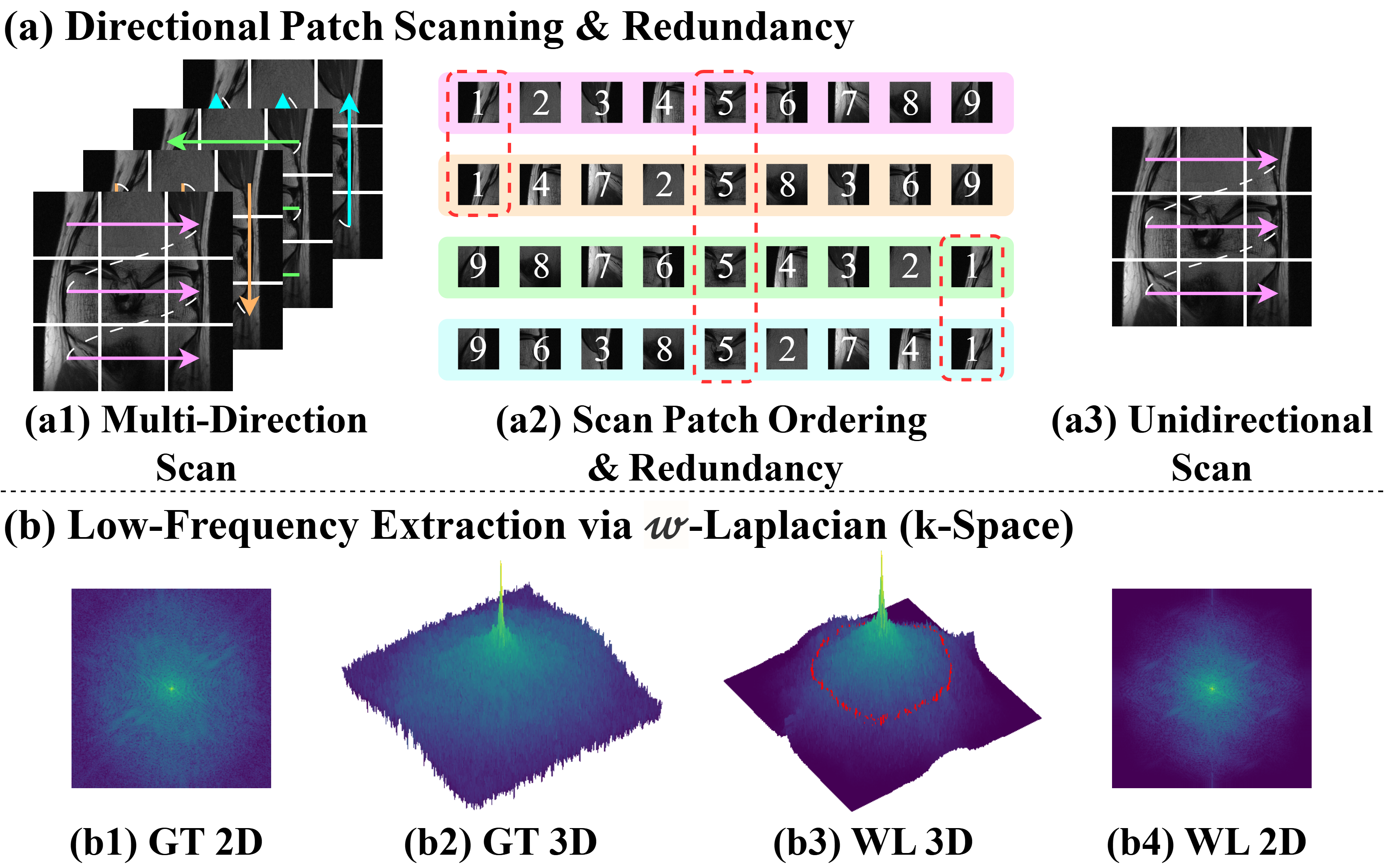}  
\caption{Illustration of scanning and decoupling.
(a) Scanning strategies. Multi-directional scanning introduces redundancy, while the unidirectional approach avoids repeated access. Colors denote scan orders; red dashed boxes highlight redundant regions.
(b) Visualization of k-space before and after the $\mathcal{w}$-Laplacian decomposition. Subfigures (b3) and (b4) show only the output branch retained for Mamba. The red circle marks the theoretical boundary between low- and high-frequency regions in k-space. This retained branch exhibits a cleaner, concentrated low-frequency spectrum and is better aligned with Mamba’s global modeling needs.}
\label{fig1}
\end{figure}

\section{Introduction}
Magnetic Resonance Imaging (MRI) is a clinically indispensable modality due to its non-invasive nature and excellent soft-tissue contrast~\cite{jerban2020mri,varela2017motion}. 
However, a primary limitation lies in its long acquisition time, which can cause patient discomfort and increase the risk of motion artifacts~\cite{hammernik2018variational,knoll2020survey}. 
To accelerate scans, modern protocols commonly adopt undersampling in the frequency domain. 
While this reduces scan time, it violates the Nyquist-Shannon sampling criterion~\cite{ye2019compressed,zbontar2018fastmri}, 
leading to aliasing artifacts and degraded image quality. Addressing this challenge requires advanced reconstruction methods capable of recovering high-fidelity images from incomplete k-space data. 
Traditional compressed sensing (CS) methods~\cite{lustig2007sparse} exploit sparsity priors in transform domains (e.g., wavelets), but they require extensive hyperparameter tuning and often lack robustness to variations in sampling patterns or noise conditions.

Recent deep learning frameworks have substantially advanced MRI reconstruction by leveraging data-driven priors from large-scale datasets~\cite{zhou2018unet++}.
Convolutional neural networks (CNNs) have demonstrated strong performance by modeling hierarchical and localized anatomical features~\cite{schlemper2017deepcascade,qin2019crnn, chen2024hi}. 
Model-based CNNs further improve reconstruction quality by integrating the MRI forward model and enforcing data consistency~\cite{hammernik2018variational,aggarwal2019modl,shan2024b0inhomogeneity}. However, the inherent locality of CNNs limits their capacity to capture long-range dependencies, which are crucial for reconstructing global anatomical structures, especially under highly undersampled conditions. 

Transformer-based architectures model global dependencies through self-attention by computing pairwise correlations across all spatial tokens~\cite{vaswani2017attention, fu2025mogo}.
This capacity has shown promise in MRI reconstruction~\cite{guo2023reconformer,wu2023transformer} by enabling global context modeling to restore anatomical structures.
Nonetheless, standard Transformers incur quadratic complexity, posing challenges for high-resolution MRI applications.
To improve efficiency, variants such as the Swin Transformer~\cite{liu2021swin} employ hierarchical designs and shifted window-based attention to restrict computations to local neighborhoods.
While computationally efficient, these approaches inherently constrain the receptive field, reigniting the fundamental trade-off between computational efficiency and global modeling capability, leaving a critical gap for a more holistic solution.

Structured State Space Models (SSMs)~\cite{gu2022efficiently} have recently emerged as promising alternatives to self-attention for global context modeling, offering linear-time scalability with strong sequence modeling capacity. Among them, Mamba~\cite{gu2023mamba} employs input-dependent state transitions and efficient gating mechanisms, enabling expressive long-range interactions with reduced computational complexity. While initially proposed for language tasks, Mamba is rapidly emerging as a powerful alternative in visual domains such as image classification and restoration~\cite{guo2024mambair,liu2024vmamba}.

However, a direct application of Mamba to MRI reconstruction reveals three fundamental limitations rooted in its original design.
First, existing vision-specific Mamba architectures utilize multi-directional spatial scanning~\cite{shi2024mambasurvey,zhang2024era} to enhance coverage, but this introduces significant computational redundancy (see Figure~\ref{fig1}(a)).

Second, its state-space parameters are derived independently for each spatial token through local transformations, limiting spatial awareness.
This design neglects neighboring context, which is essential for modeling coherent anatomical structures in MRI.
Third, existing Mamba-based variants lack sensitivity to high-frequency components~\cite{ma2024tinyvim}, which are critical for preserving fine anatomical details in MRI reconstruction.

To address these limitations, we propose HiFi-Mamba (short for \textbf{Hi}gh-\textbf{Fi}delity Mamba), a novel reconstruction framework built upon an efficient, dual-stream Mamba-based architecture including:

\begin{itemize}

    \item \textit{A novel HiFi-Mamba block} that embodies an efficient, dual-stream architecture. It employs a unidirectional scan for efficiency and a cross-stream guidance mechanism to resolve the locality and high-frequency insensitivity inherent to Mamba.

    \item \textit{A lightweight $\mathcal{w}$-Laplacian block} that decomposes features into high- and low-frequency streams, enabling frequency-aware dual-stream processing in our HiFi-Mamba.
    
    \item \textit{State-of-the-art results}. On public MRI benchmarks, HiFi-Mamba consistently outperforms leading CNN-, Transformer-, and Mamba-based baselines, while establishing a superior trade-off between reconstruction fidelity and computational efficiency.

\end{itemize}



    


\section{Related Work}

\paragraph{CNN-based MRI Reconstruction.}
CNNs have been widely employed in MRI reconstruction for their ability to extract hierarchical features efficiently.
Early models such as DeepCascade~\cite{schlemper2017deepcascade} and ISTA-Net~\cite{zhang2018ista} framed reconstruction as an unrolled optimization process, integrating data fidelity with learnable, network-based priors.
Subsequent approaches, including KIKI-Net~\cite{eo2018kikinet} and DuDoRNet~\cite{zhou2020dudornet}, incorporated architectural advances such as residual connections, dilated convolutions, and hybrid modeling across image and k-space domains to enhance reconstruction quality and optimization stability.
However, the inherently localized receptive fields of CNNs constrain their ability to capture global anatomical context. This limitation has motivated the development of architectures with enhanced global modeling capacity.

\begin{figure*}[htbp]
    \centering
    \includegraphics[width=0.99\linewidth]{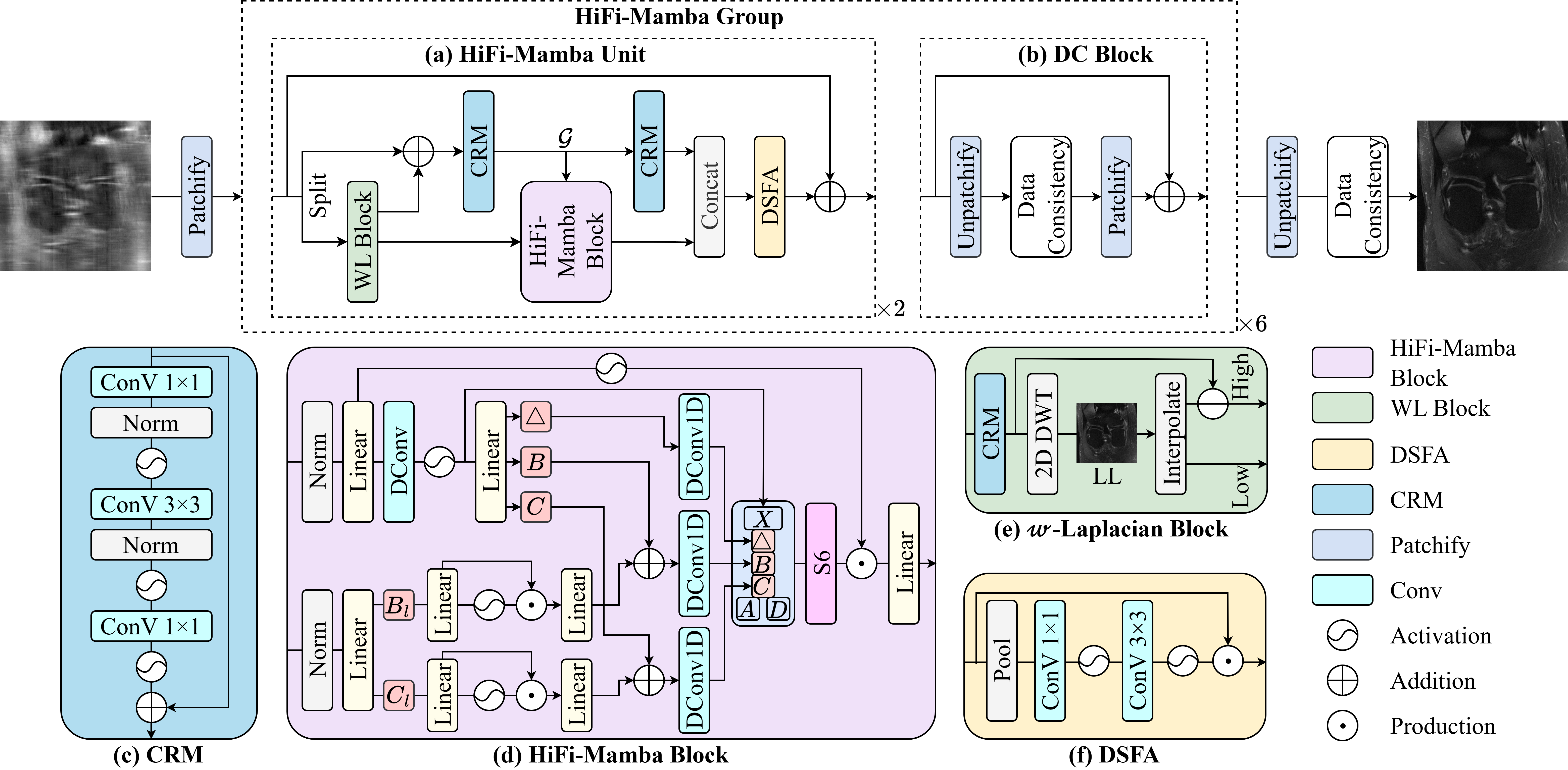}
    \caption{Overview of the proposed HiFi-Mamba architecture. (a) The HiFi-Mamba Unit splits the input into high- and low-frequency components via the $\mathcal{w}$-Laplacian Block, processes them using the HiFi-Mamba Block and CRM (Condition Refinement Module), and fuses them with DSFA (Dual-Stream Fusion Attention). (b) The data consistency block. (c) CRM performs cross-resolution feature transformation. (d) The HiFi-Mamba block models frequency-aware sequences using Mamba-based token mixing. (e) The $\mathcal{w}$-Laplacian block performs idelity-preserving spectral decoupling. (f) DSFA fuses dual-frequency streams with adaptive weighting.}
    \label{fig:overview}
\end{figure*}

\paragraph{Transformer-based MRI Reconstruction.}
Transformers have been increasingly adopted in MRI reconstruction for their ability to model long-range dependencies through global self-attention, which facilitates the preservation of anatomical structures.
Early models such as SLATER~\cite{you2021unsupervised} and  TTM~\cite{gu2021texturetransformer} applied Transformer blocks in the image domain.
DuDReTLU-Net~\cite{wang2023dual} jointly models image and $k$-space domains through a transformer-based architecture to enhance reconstruction quality.
Later variants such as SwinMR~\cite{yun2023swinmr} and ReconFormer~\cite{guo2023reconformer} adopted hierarchical and windowed self-attention mechanisms to improve multi-scale feature modeling and computational efficiency.
Although these designs reduce computational overhead, they often rely on localized attention and staged aggregation, which may still limit the capacity for global context modeling in high-resolution MRI reconstruction.

\paragraph{Structured State Space Models and Mamba.}
State space models (SSMs) have recently gained attention for their ability to model long-range dependencies with linear complexity~\cite{li2024survey}. The vision-centric Mamba variant, vMamba~\cite{liu2024vmamba}, introduced a four-directional scanning strategy to enhance 2D spatial context modeling. Recent works, such as MambaMIR~\cite{zhao2024mambamir} and LMO~\cite{li2025lmo}, have extended this paradigm for MRI reconstruction.
However, these adaptations remain suboptimal in both efficiency and fidelity. Their reliance on multi-directional scanning induces considerable computational overhead due to repeated processing of spatial features—an issue exacerbated in high-resolution scenarios. Moreover, they inherit two core limitations of the original Mamba design: purely local state-space parameterization and insensitivity to high-frequency anatomical details. These constraints highlight a gap between current vision-based Mamba variants and the spectral characteristics of MRI.

\subsection{Overall Pipeline}
Our proposed HiFi-Mamba network follows an unrolled optimization framework, a powerful paradigm for solving inverse problems like MRI reconstruction. The network backbone consists of $K=8$ cascaded HiFi-Mamba Groups. As shown in Figure~\ref{fig:overview}, each Group consists of two sequential Mamba Units for feature refinement, followed by a Data Consistency (DC) block.
Specifically, the input undersampled image $I_{\text{in}} \in \mathbb{R}^{H \times W \times 2}$ is a two-channel tensor representing the real and imaginary parts of the complex-valued MR data. 
The input tensor is first transformed into patch embeddings as:
\begin{equation}
F_{\text{in}} = \mathcal{P} (I_{\text{in}}), \quad F_{\text{in}} \in \mathbb{R}^{H/P \times W/P \times C}.
\end{equation}
where $\mathcal{P}(\cdot)$ denotes the patch embedding operation.
The patch embeddings $F_{\text{in}}$ are then processed sequentially through the $K$ Groups. Within each Group, the two Mamba Units progressively refine the features by modeling both global and local dependencies via a novel asymmetric dual-stream architecture. The subsequent DC block then enforces data fidelity by incorporating the originally acquired k-space measurements.

After the final Group and its subsequent DC block, an unpatchifying layer restores the features to the full image resolution $\mathbb{R}^{H \times W \times 2}$. Finally, a DC block is applied in the image domain to ensure global data fidelity before producing the output reconstruction $I_{\text{out}} \in \mathbb{R}^{H \times W \times 2}$.

\subsection{The Mamba Unit}
The Mamba Unit is the core module in each HiFi-Mamba Group, following a dual-stream, frequency-aware architecture. As shown in Figure~\ref{fig:overview}(a), it adopts an asymmetric structure to adaptively process MRI-specific features.

\paragraph{Stream Preparation via Frequency Decoupling.}
The unit's workflow begins by splitting the input feature map $F_\text{in} \in \mathbb{R}^{H \times W \times C}$ into two feature maps, $F_1, F_2 \in \mathbb{R}^{H \times W \times \frac{C}{2}}$. The feature map $F_1$ is processed by the lightweight $\mathcal{w}$-Laplacian (WL) block to yield a low-frequency component $F_{\text{low}}$ and a residual high-frequency component $F'_{\text{high}}$, i.e.,
\begin{equation}
 F_{\text{low}}, F'_{\text{high}} = \mathrm{WL}(F_1).
\end{equation}
The second feature map, $F_2$, is then fused with $F'_{\text{high}}$ via element-wise addition to form the final high-frequency stream for parallel processing, i.e.,
\begin{equation}
 F_{\text{high}} = F_2 + F'_{\text{high}}.
\end{equation}

\paragraph{Asymmetric Parallel Processing.}
The high-frequency feature map $F_{\text{high}}$ is first processed by a dedicated Condition Refinement Module (CRM; see Figure~\ref{fig:overview}(c)) to extract an anatomical guidance feature:
\begin{equation}
\mathcal{G} = \operatorname{CRM}(F_{\text{high}}). \label{eq:crm1}
\end{equation}
Here, $\mathcal{G}$ captures spatial high-frequency structures and serves as a guidance prior for modulating the low-frequency stream.
To further enhance the high-frequency representation, $\mathcal{G}$ is subsequently refined by a second CRM:
\begin{equation}
\tilde{F}_{\text{high}} = \operatorname{CRM}(\mathcal{G}). \label{eq:crm2}
\end{equation}
Concurrently, the low-frequency feature map $F_{\text{low}}$ is processed by our novel HiFi-Mamba Block. We denote this operation as $\mathcal{H}(\cdot)$, which performs long-range dependency modeling under the explicit guidance of the anatomical map $\mathcal{G}$, i.e.,
\begin{equation}
 \tilde{F}_{\text{low}} = \mathcal{H}(F_{\text{low}} \mid \mathcal{G}).
\end{equation}
This cross-stream guidance, detailed in the HiFi-Mamba Block section, infuses essential high-frequency cues into the global context modeling.

\paragraph{Dual-Stream Fusion.}
The two enhanced feature maps, $\tilde{F}_{\text{low}}$ and $\tilde{F}_{\text{high}}$, are concatenated and fused by a Dual-Stream Fusion Attention (DSFA) module (Figure~\ref{fig:overview}(f)) to a fused feature map, $F_{\mathrm{fused}}$:
\begin{equation}
 F_{\rm fused} = \operatorname{DSFA}\left(\operatorname{concat}([\tilde{F}_{\text{low}}, \tilde{F}_{\text{high}}])\right).
\end{equation}
The final output of the unit, $F_{\text{out}}$, is then obtained by applying a residual connection with the unit's original input, $F_\text{in}$:
\begin{equation}
 F_{\text{out}} = F_{\rm fused} + F_\text{in}.
\end{equation}

\subsection{$\mathcal{w}$-Laplacian Block}

To enable frequency-aware dual-stream processing, we implement the $\mathcal{w}$-Laplacian block (Figure~\ref{fig:overview}(e)) to perform fidelity-preserving spectral decoupling. This operation serves two purposes: (1) providing a dedicated low-frequency input for the Mamba stream (see Figure~\ref{fig1}(b)), and (2) extracting high-frequency features into a parallel stream for fine-detail enhancement.

While the Laplacian pyramid~\cite{burt1987laplacian} offers residual-based multiscale representations, its decomposition is resolution-oriented rather than frequency-structured. The resulting low-frequency component is a blurred, downsampled approximation without explicit spectral semantics, limiting its utility for preserving anatomical context in MRI. To overcome this limitation, we replace the resolution-based hierarchy with a wavelet-based formulation that enables structured and reversible frequency separation.

The $\mathcal{w}$-Laplacian block begins by refining the input feature map $F_1 \in \mathbb{R}^{H \times W \times \frac{C}{2}}$ using a CRM to enhance local feature representation, yielding:
\begin{equation}
    F'_{1} = \operatorname{CRM}(F_1).
\end{equation}

A channel-wise 2D discrete wavelet transform (DWT) is then applied to $F'_{1}$ to obtain the four standard subbands:
\begin{equation}
    \text{DWT}(F'_{1})=\{LL, LH, HL, HH\}.
\end{equation}

To avoid potential information loss and basis dependency when using the high-frequency subbands directly, a smoothed low-frequency base is first formed by upsampling the $LL$ subband, i.e.,
\begin{equation}
    F_{\text{low}} = \text{Upsample}(LL).
\end{equation}
The complementary high-frequency component is then defined as the residual between the refined map and this base:
\begin{equation}
    F_{\text{high}} = F'_{1} - F_{\text{low}}.
\end{equation}

This decoupling enables our specialized dual-stream processing: the low-frequency stream routes $F_{\text{low}}$ to the HiFi-Mamba block for long-range anatomical modeling, while the high-frequency stream processes $F_{\text{high}}$ for targeted enhancement.

\subsection{HiFi-Mamba Block}
\label{sec:hifimamba_block}
As a central component of our dual-stream architecture, the HiFi-Mamba block processes the low-frequency feature map, $F_{\text{low}}$, conditioned by the anatomical guidance map, $\mathcal{G}$. This design directly addresses the core limitations of standard Mamba for MRI reconstruction by introducing two key modifications: a cross-frequency guidance mechanism and a spatially-aware parameter refinement process. The block's structure is illustrated in Figure~\ref{fig:overview}(d).

\paragraph{Initial Parameter Generation.}
The block first produces initial state-space parameters from the low-frequency feature map $F_{\text{low}}$. To embed local spatial context prior to the main selective scan, this map is projected and split into a main feature map, $F_c$, and a residual map, $Z$:
\begin{equation}
    F_c, Z = \operatorname{Split}(\operatorname{Linear}(\operatorname{Norm}(F_{\text{low}}))).
\end{equation}
The main feature map $F_c$ is then passed through a 2D convolution and a SiLU activation to produce a context-aware feature map, $F_{\text{conv}}$, i.e.,
\begin{equation}
    F_{\text{conv}} = \operatorname{SiLU}(\operatorname{Conv2D}(F_c)).
\end{equation}
After reshaping $F_{\text{conv}}$ into a sequence $F_s$, a subsequent linear projection is applied, followed by a split operation, to generate the initial state projection matrices $B, C \in \mathbb{R}^{B \times d_s \times L}$ and the timestep parameter $\Delta \in \mathbb{R}^{B \times d_t \times L}$, i.e.,
\begin{equation}
    [\Delta,\,B,\,C] = \operatorname{Split}(\operatorname{Linear}(F_s)).
\end{equation}
At this stage, while containing some local context from the convolution, these parameters have not yet been informed by the high-frequency stream.



\paragraph{Cross-Frequency Guidance.}
To address the first core limitation—Mamba's insensitivity to high frequencies—conditioning terms are derived from the anatomical guidance map $\mathcal{G}$. An initial projection is applied to $\mathcal{G}$ to produce pre-conditioned tensors $B'_{h}$ and $C'_{h}$. These tensors are then processed by two independent gating mechanisms to generate the final conditioning terms $B_h$ and $C_h$. The process for generating $B_h$ from $B'_{h}$ is as follows:
\begin{equation}
\begin{aligned}
    B'_{h,1}, B'_{h,2} &= \operatorname{Split}(\operatorname{Linear}(B'_{h})), \\
    B_h &= \operatorname{GELU}(B'_{h,1}) \odot B'_{h,2}.
\end{aligned}
\end{equation}
The term $C_h$ is generated from $C'_{h}$ through an identical process but with a separate set of learned weights. These conditioning terms are subsequently integrated through element-wise addition:
\begin{equation}
    B = B + B_h, \qquad C = C + C_h.
\end{equation}
The efficacy of this mechanism lies in its targeted modulation of the two fundamental processes of the SSM, governed by the state and output equations:
\begin{equation}
    h'(t) = Ah(t) + Bx(t), \quad 
    y(t) = Ch(t).
\end{equation}
Here, the gating mechanisms act as dynamic filters, selectively distilling salient high-frequency information (e.g., anatomical edges, aliasing artifacts) from $\mathcal{G}$ into the conditioning terms. The conditioning term $B_h$ modulates the input projection matrix $B$, thereby conditioning the influence of the input signal $x(t)$ on the state vector $h(t)$ based on critical local details. Simultaneously, the term $C_h$ modulates the output projection matrix $C$, which determines the projection from the state to the output, ensuring that these distilled high-frequency details are accurately rendered in the final reconstruction.

\paragraph{Spatially-Aware Refinement.}
To address the second limitation—the strictly local parameter generation in standard Mamba—the conditioned state projection matrices ($B, C$) and the timestep parameter ($\Delta$) are refined. Each is processed by a dedicated 1D depth-wise convolution with a kernel size of $k=7$. 
This step allows the parameters for each token to be influenced by its spatial neighbors, injecting essential local context. Applying convolution to $\Delta$ additionally ensures that the state transition dynamics evolve smoothly across the spatial sequence.

\paragraph{Output Generation.}
The refined parameters ($\Delta, B, C$) are then supplied to the selective scan operation to produce an output feature map, $F_{ssm}$. The final output of the block, $\tilde{F}_{\text{low}}$, is obtained by modulating $F_{ssm}$ with the residual feature map $Z$ and applying a final linear projection, i.e.,
\begin{equation}
    \tilde{F}_{\text{low}} = \operatorname{Linear_{out}}(F_{ssm} \odot \operatorname{SiLU}(Z)).
\end{equation}
This final operation combines the long-range context from $F_{ssm}$ with local features from $Z$ through a gated modulation, followed by a final linear projection.

\begin{table*}[t]
\centering
\scriptsize
\resizebox{\textwidth}{!}{%
\begin{tabular}{@{}l|cc|cc|cc|cc|cc|cc@{}}
\hline
\multirow{3}{*}{\textbf{Method}} 
& \multicolumn{6}{c|}{\textbf{fastMRI}} 
& \multicolumn{6}{c}{\textbf{CC359}} \\
\cline{2-13}
& \multicolumn{2}{c|}{PSNR$\uparrow$} 
& \multicolumn{2}{c|}{SSIM$\uparrow$} 
& \multicolumn{2}{c|}{NMSE$\downarrow$} 
& \multicolumn{2}{c|}{PSNR$\uparrow$} 
& \multicolumn{2}{c|}{SSIM$\uparrow$} 
& \multicolumn{2}{c}{NMSE$\downarrow$} \\
\cline{2-13}
& 4$\times$ & 8$\times$ & 4$\times$ & 8$\times$ & 4$\times$ & 8$\times$
& 4$\times$ & 8$\times$ & 4$\times$ & 8$\times$ & 4$\times$ & 8$\times$ \\
\hline
Zero-Filling & 29.25 & 25.95 & 0.723 & 0.620 & 0.035 & 0.064 & 24.79 & 21.27 & 0.725 & 0.576 & 0.053 & 0.120 \\
\hline
UNet~\cite{zhou2018unet++} & 31.66 & 28.60 & 0.798 & 0.697 & 0.021 & 0.035 & 28.27 & 24.28 & 0.847 & 0.720 & 0.025 & 0.059 \\
ISTA~\cite{zhang2018ista} & 33.27 & 29.44 & 0.832 & 0.714 & 0.012 & 0.030 & 32.03 & 25.44 & 0.902 & 0.744 & 0.010 & 0.046 \\
\hline
ReconFormer~\cite{guo2023reconformer} & 33.75 & 30.42 & 0.837 & 0.728 & 0.011 & 0.026 & 32.46 & 26.47 & 0.906 & 0.766 & 0.010 & 0.039 \\
FpsFormer~\cite{meng2025boosting} & 33.74 & 30.63 & 0.841 & 0.732 & 0.011 & 0.026 & 32.35 & 26.65 & 0.897 & 0.768 & 0.010 & 0.038 \\
\hline
LMO~\cite{li2025lmo} & 34.49 & 31.10 & 0.846 & 0.744 & 0.011 & 0.023 & 35.35 & 27.99 & 0.921 & 0.787 & 0.006 & 0.028 \\
HiFi-Mamba (P2) & 34.47 & 31.38 & 0.853 & 0.758 & 0.010 & 0.021 & 35.74 & 28.08 & 0.935 & 0.802 & 0.005 & 0.027 \\
\textbf{HiFi-Mamba (P1)} & \textbf{34.85} & \textbf{31.81} & \textbf{0.855} & \textbf{0.762} & \textbf{0.010} & \textbf{0.020} & \textbf{36.93} & \textbf{28.49} & \textbf{0.942} & \textbf{0.810} & \textbf{0.004} & \textbf{0.026} \\
\hline
\end{tabular}%
}
\caption{Quantitative results on fastMRI-Equispaced and CC359-Equispaced under $4\times$ and $8\times$ accelerations.}
\label{tab:mri_comparison}
\end{table*}

\section{Experiments}

\subsection{Experimental Settings}
\subsubsection{Datasets and Evaluation Metrics.}
We evaluate HiFi-Mamba on two public MRI datasets: fastMRI (knee)~\cite{zbontar2018fastmri} and CC359 (brain)~\cite{warfield2004staple}.
The fastMRI dataset comprises 1,172 complex-valued single-coil coronal knee scans acquired with Proton Density (PD) weighting. Each scan contains approximately 35 coronal slices with a matrix size of $320\times320$. We exclusively use the Proton Density Fat-Suppressed (PDFS) subset for both training and evaluation, following the official data split.
The CC359 dataset contains raw brain MR scans acquired using clinical MR scanners (Discovery MR750; GE Healthcare, USA). Following the official split, 25 subjects are used for training and 10 for testing. Each slice has a matrix size of $256\times256$.
To eliminate slices with limited anatomical content, we discard the first and last five slices for fastMRI and the first and last 15 slices for CC359.

In our experiments, the inputs are generated by applying equispaced 1D Cartesian undersampling masks, as provided by the fastMRI challenge~\cite{zbontar2018fastmri}. Specifically, for an acceleration factor (AF) of 4, the central 8\% of k-space lines are fully sampled; for $\text{AF}=8$, this portion is reduced to 4\%.


We evaluate reconstruction performance using three widely used metrics: PSNR (Peak Signal-to-Noise Ratio)~\cite{huynh2008scope}, SSIM (Structural Similarity Index)~\cite{wang2004image}, and NMSE (Normalized Mean Squared Error)~\cite{zhao2016loss}.

\subsubsection{Data Preprocessing Strategy.}
Previous MRI reconstruction studies employ diverse preprocessing strategies (e.g., normalization schemes), leading to inconsistencies in intensity distributions that hinder fair comparisons.
To mitigate this, we adopt a unified and transparent preprocessing strategy, applied consistently across experiments. This ensures reproducibility and minimizes confounding variables during inter-method comparison. Detailed pipeline specifications (e.g., normalization, k-space undersampling) are provided in the appendix to support reproducibility and future benchmarking.


\subsubsection{Training Details.}
Our model consists of a stacked 6$\times$2 configuration of HiFi-Mamba units. 
We use the AdamW optimizer with an initial learning rate of $1\!\times\!10^{-3}$. A cosine annealing schedule with a 5-epoch warm-up is used, and training is performed for 100 epochs with a batch size of 4. An $\ell_1$ loss is adopted for MRI reconstruction.

All experiments are conducted on two NVIDIA H100 GPUs. FLOPs are measured on a single NVIDIA A100 GPU. The implementation is based on PyTorch.

%
%


\subsection{Comparison with State-of-the-Art Methods}

We compare HiFi-Mamba against representative state-of-the-art methods spanning three major model paradigms: CNN-based (UNet, ISTA-Net), Transformer-based (ReconFormer, FPSFormer), and Mamba-based (LMO). Evaluations are performed using equispaced 1D Cartesian undersampling at acceleration factors (AF) of 4× and 8× on the fastMRI and CC359 datasets. The quantitative results are summarized in Table~\ref{tab:mri_comparison}.

HiFi-Mamba (P2), where P2 denotes a patch size of 2, consistently outperforms existing methods across most evaluation scenarios. Notably, it achieves significant gains at $\text{AF}=8$ on both datasets, e.g., reaching 31.38 dB PSNR and 0.758 SSIM on fastMRI, and 28.08 dB PSNR and 0.802 SSIM on CC359. Although its PSNR (34.47 dB) on fastMRI at $\text{AF}=4$ is slightly lower than LMO (34.49 dB), HiFi-Mamba (P2) still surpasses LMO in SSIM and NMSE and consistently outperforms all other methods across the remaining settings, demonstrating strong generalization capability.

\begin{table}[t]
\centering
\scriptsize

\resizebox{\columnwidth}{!}{%
\begin{tabular}{@{}lcccc@{}}
\hline
\textbf{Method} & \textbf{Scanning} & \textbf{FLOPs} & \textbf{PSNR}$\uparrow$ & \textbf{SSIM}$\uparrow$ \\
\hline
ReconFormer & Attention & 270.60G & 30.42 & 0.728 \\
FpsFormer & Attention & 200.45G & 30.63 & 0.732 \\
LMO & 4-Directional & 484.98G & 31.10 & 0.744 \\
HiFi-Mamba (P1) & 1-Directional & 270.37G & \textbf{31.81} & \textbf{0.762} \\
\textbf{HiFi-Mamba (P2)} & 1-Directional & \textbf{67.87G} & 31.38 & 0.758 \\
\hline
\end{tabular}%
}
\caption{Efficiency comparison on fastMRI ($\text{AF}=8$, $320\times320$) using NVIDIA A100.}
\label{tab:efficiency_cc359}
\end{table}

The finer-grained variant HiFi-Mamba (P1), with a patch size of 1, further improves reconstruction performance, establishing a new state-of-the-art. Specifically, it achieves $34.85$ / $31.81$ dB PSNR and $0.855$ / $0.762$ SSIM on fastMRI ($\text{AF}=4$ / $8$), with corresponding NMSE of $0.010$ / $0.020$. On CC359, it achieves 36.93 dB PSNR and 0.942 SSIM at $\text{AF}=4$, and maintains robust performance at $\text{AF}=8$, reaching 28.49 dB PSNR, 0.810 SSIM, and 0.026 NMSE, consistently surpassing prior state-of-the-art methods across all metrics.

Collectively, these results highlight the effectiveness of our frequency-aware architecture and fine-grained feature modeling in achieving robust, high-fidelity MRI reconstruction under aggressive acceleration.


\begin{figure*}[htbp]
    \centering
    \includegraphics[width=\linewidth]{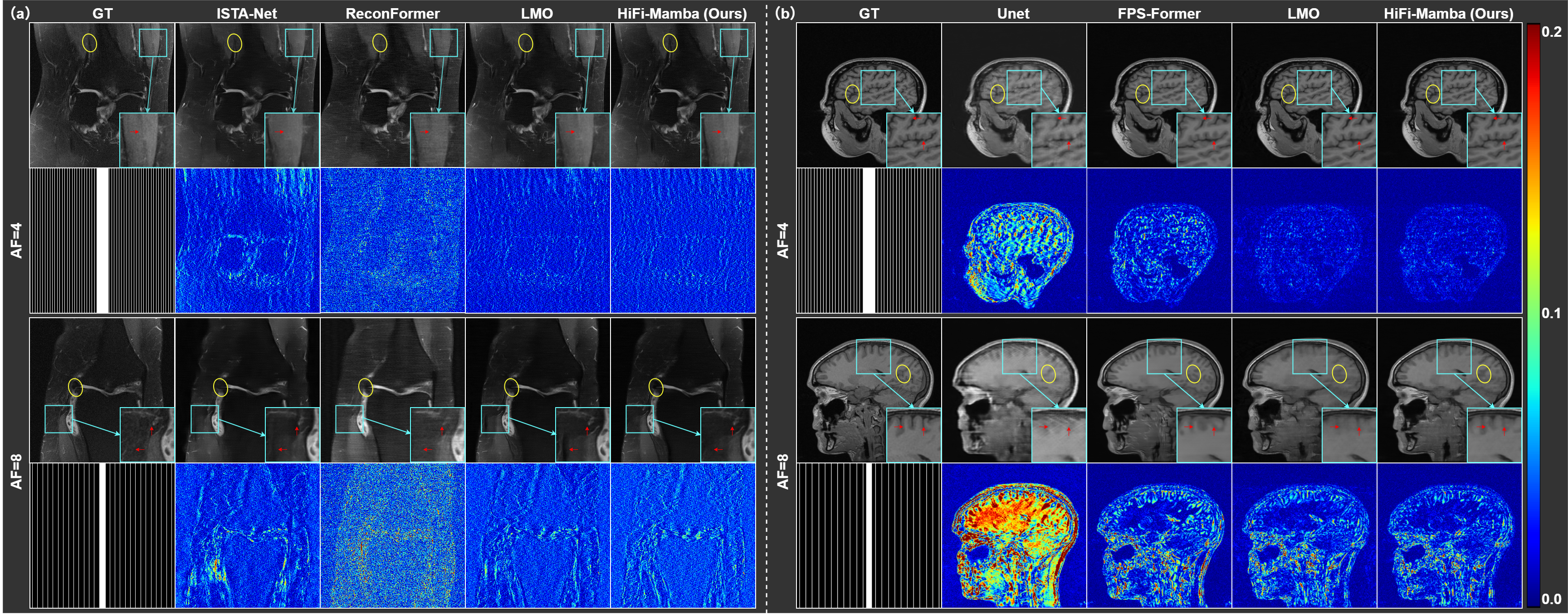}
    \caption{Qualitative comparison on the fastMRI and CC359 datasets under single-coil settings. (a) Reconstruction results on the fastMRI knee dataset with acceleration factors $\text{AF}=4$ and $\text{AF}=8$. (b) Reconstruction results on the CC359 brain dataset under the same acceleration factors. The second row of each subplot shows the corresponding error maps. The blue boxes, yellow ellipses and red arrow highlight the details in the reconstruction results.}
    \label{fig:result}
\end{figure*}
\subsubsection{Visualization Results.}

Figure~\ref{fig:result} presents qualitative comparisons under $4\times$ and $8\times$ acceleration on representative slices from the fastMRI (knee) and CC359 (brain) datasets. The top row shows the reconstructed images, while the bottom row displays the corresponding error maps (difference from ground truth), color-coded from 0 (blue) to 0.2 (red). Yellow circles and red arrows indicate discrepancies in structural detail, while blue boxes mark zoomed-in regions for closer inspection.

ISTA-Net and UNet exhibit edge blurring and loss of structural detail. ReconFormer and FPS-Former partially alleviate these issues but still fail to reconstruct fine anatomical features with high acceleration factors. Notably, FPS-Former produces visually sharp contours in $\text{AF}=8$ brain reconstructions, but introduces hallucinated structures inconsistent with the ground truth. The Mamba-based LMO also exhibits boundary degradation and missing details.

In contrast, HiFi-Mamba demonstrates robustness to various anatomical structures and acceleration factors.  These advantages are also reflected in the corresponding error maps. 


\subsubsection{Efficiency}
We evaluate computational efficiency on the fastMRI dataset ($\text{AF} = 8$, $320\times320$) using a single NVIDIA A100 GPU. 
As shown in Table~\ref{tab:efficiency_cc359}, HiFi-Mamba achieves a favorable balance between performance and efficiency.
HiFi-Mamba (P2) achieves 31.38\,dB PSNR and 0.758 SSIM with only 67.87G FLOPs—substantially lower than all competing models, while matching or surpassing heavier attention-based architectures such as ReconFormer and FpsFormer. 
HiFi-Mamba (P1), though requiring 270.37G FLOPs, delivers the best overall performance (31.81\,dB PSNR, 0.762 SSIM), outperforming the state-of-the-art LMO (484.98G FLOPs) in both accuracy and efficiency. 
These results demonstrate the resource-awareness of our design for high-fidelity MRI reconstruction.

\subsection{Ablation Studies and Analysis}
\label{sec:ablation}

\begin{table}[t]
\footnotesize
\centering

\resizebox{\columnwidth}{!}{%
\begin{tabular}{@{}ccccccc@{}}
\hline
$\mathcal{w}$-Lap & HiFi-Mamba & DSFA & CRM & PSNR$\uparrow$ & SSIM$\uparrow$ & NMSE$\downarrow$ \\
\hline
\checkmark &  &  &  & 27.07 & 0.790 & 0.032 \\
\checkmark & \checkmark &  &  & 27.46 & 0.794 & 0.030 \\
\checkmark & \checkmark & \checkmark &  & 27.99 & 0.799 & 0.028 \\
\checkmark & \checkmark & \checkmark & \checkmark & \textbf{28.07} & \textbf{0.802} & \textbf{0.027} \\
\hline
\end{tabular}%
}
\caption{Ablations on CC359 ($\text{AF}=8$, $\text{patch}=2$).}
\label{tab:cc359_ablation}
\end{table}

\subsubsection{Ablation on Mamba Unit.}

We perform an ablation study on the CC359 dataset with $\text{AF}=8$ to assess the contribution of each component. Starting with the proposed $\mathcal{w}$-Laplacian transform, we obtain 27.07 dB PSNR and 0.790 SSIM. Incorporating the HiFi-Mamba block, which enables frequency-aware interaction, increases the PSNR to 27.46 dB and the SSIM to $0.794$.
Adding the DFSA module refines the fused features, improving performance to 27.99 dB PSNR and 0.799 SSIM. Finally, appending the CRM module enhances high-frequency stream representations and achieves the best performance: 28.07 dB PSNR, 0.802 SSIM, and 0.027 NMSE. These results demonstrate the complementary benefits of each module and their collective contribution to reconstruction quality.

\begin{table}[t]
\footnotesize
\centering

\begin{tabular}{cccccc}
\hline
\textbf{Patch-Size} & \textbf{Depth}  & \textbf{PSNR} & \textbf{SSIM} & \textbf{NMSE} \\
\hline
2&      $3\times2$  & 27.92 & 0.800 & 0.028 \\
2 &  $4\times2$ & 27.95 & 0.800 & 0.028 \\
2 &     $6\times2$    & 28.07 & 0.802 & 0.027 \\
2 &     $8\times2$    & 28.15 & 0.805 & 0.027 \\
4 &  $6\times2$ & 27.71 & 0.793 & 0.029 \\
1 &   $6\times2$ & \textbf{28.49} & \textbf{0.810} & \textbf{0.026} \\
\hline
\end{tabular}

\caption{Model architecture experiment is conducted on the CC359 dataset with $\text{AF}=8$.}
\label{tab:cc359_ablation_depth}
\end{table}

\paragraph{Ablation on Model Depth and Patch Size.}
We perform an ablation study on model depth and patch size under an $8\times$ acceleration setting using the CC359 dataset. As shown in Table~\ref{tab:cc359_ablation_depth}, with patch size fixed at 2, increasing depth from $3\times2$ to $6\times2$ consistently improves PSNR and SSIM, indicating better anatomical modeling. Further increasing to $8\times2$ offers only marginal gains, suggesting saturation beyond moderate depth.

Conversely, with depth fixed at $6\times2$, decreasing patch size from 4 to 1 steadily improves all metrics. The best performance is achieved with patch size 1 and depth $6\times2$, reaching 28.49 dB PSNR, 0.810 SSIM, and 0.026 NMSE.

These results confirm the scalability of our design—deeper models and finer spatial granularity enhance reconstruction without overfitting, highlighting the robustness and flexibility of the proposed architecture.

\paragraph{Ablation on Kernel Size.}
We evaluate the impact of kernel size in the spatially-aware refinement module by varying the receptive field of the depthwise 1D convolution applied to the conditioned parameters ($B$, $C$, $\Delta$). As shown in Table~\ref{tab:cc359_ablation_size}, increasing the kernel size from $3$ to $7$ yields consistent improvements across all metrics. The kernel size 7 configuration achieves the best performance, indicating that larger spatial context enhances the anatomical coherence of the learned dynamics.

\begin{table}[t]
\footnotesize
\centering

\begin{tabular}{cccc}
\hline
\textbf{Mechanism} & \textbf{PSNR} & \textbf{SSIM} & \textbf{NMSE} \\
\hline
 DConv1D ($k= 3$) & 27.81 & 0.796 & 0.030 \\
 DConv1D ($k=5$) & 28.05 & 0.805 & 0.028 \\
 DConv1D ($k=7$) & 28.49 & 0.810 & 0.026 \\
\hline
\end{tabular}

\caption{Spatially-Aware experiment is conducted on the CC359 dataset with $\text{AF}=8$, and $\text{patch size} =1$. Here, $k$ denotes the kernel size.}
\label{tab:cc359_ablation_size}
\end{table}

\section{Conclusion}
In this paper, we present HiFi-Mamba, a frequency-aware
dual-stream architecture for MRI reconstruction. By coupling a $\mathcal{w}$-Laplacian block for spectral decoupling with a
guided Mamba block that models global anatomy while integrating high-frequency detail, HiFi-Mamba addresses key limitations of prior state-space models, including redundant scanning, local-only parameterization, and frequency insensitivity. Extensive experiments demonstrate that HiFi-Mamba achieves state-of-the-art reconstruction accuracy while substantially reducing computational cost. We believe it offers a promising direction for frequency-structured and efficiency-aware modeling in MRI reconstruction.
\section{Acknowledgments}
This work was supported by the National Natural Science Foundation of China under
Grant 62301352.
\bibliography{aaai2026}

\end{document}